\documentstyle[aps,prl]{revtex}
\begin{document}
\title{Macroscopic Quantum Oscillations of Antiferromagnetic Nanoclusters
in a Swept Magnetic Field }
\author{A.K. Zvezdin}
\address{Institute of General Physics, Russian Academy of
Sciences, Vavilov st..38, 117942, Moscow , Russia }
\date{\today}
\maketitle
\begin{abstract}
New macroscopic quantum interference effects in the behavior of
the antiferromagnetic nanoclusters under action of a swept
magnetic field are predicted and theoretically investigated,
namely : oscillations of the magnetic susceptibility of
nanocluster similar to the electron Bloch oscillations in crystal
and the corresponding Stark ladder-like resonances in ac-magnetic
field. These effects are related to that of the macroscopic
quantum coherence phenomena. The analogy of the antiferromagnetic
nanoclusters quantum behavior and Josephson junction with a small
capacity was revealed as well.
\end{abstract}
\vspace{0.2cm}
\vspace{0.2cm}

The quantum dynamics of magnetic nanoclusters with a
large spin is drawing increasing attention, both from experiment
and theory. Fundamental scientific questions at stake are the
macroscopic quantum phenomena including the macroscopic tunneling
and interference of wave functions and the transition from quantum
to classical behavior -- a process known as decoherence. From this
point of view metallorganic molecules with large number of d-ions
(Fe, Mn, Co etc) such as Mn$_{12}$, Fe$_{8}$, Fe$_{10}$ are of
particular interest. The presence of strong enough
antiferromagnetic interaction between d-ions is a characteristic
feature of these objects. Some of these clusters have the magnetic
structure of the ferrimagnetic type (Mn$_{12}$, Fe$_{8}$) with
large spin moment in the ground state ($S=10$). As of now the
quantum mechanical properties of these nanoclusters are
investigated in great detail. Recently such new phenomena and
specific features as molecular bistability \cite{1}, macroscopic
quantum tunneling of magnetization, quantum hysteresis
\cite{2,3,4}, the appearance of Berri--phase\cite{5} and quantum
peculiarity  of magnetic susceptibility behavior \cite{6}, have
been discovered and predicted. The quantum properties of
nanoclusters with magnetic structure of antiferromagnet type
(Fe$_{10}$, Fe$_{6}$) are less well understood \cite{7}. The aim
of the present paper is the investigation of macroscopic quantum
dynamics phenomena of the antiferromagnetic clusters in a swept
magnetic field. The reason to use the swept field is that it
produces a torque on the spin system of the antiferromagnetic
molecules accelerating the spin precession and displays a new
macroscopic quantum interference effects in their behavior.

We consider the ring-like molecular magnets (Fe$_{10}$,
Fe$_{6}$, Fe$_{18}$) composed of $N=2k$ $d$--ions spaced on a
circle lying in the $xy$-plane; there is an antiferromagnetic
exchange interaction between them. The Hamiltonian of this system
can be represented as
\begin{equation}\label{H1}
  {\mathcal{H}}=J\left(\sum_{i=1}^{N-1}\vec{S}_{i}
  \vec{S}_{i+1}+\vec{S}_{1}\vec{S}_{N}\right
  )+\sum_{i=1}^{N}H_{cr}\left(S_{i}\right)+
  g\mu_{B}\sum_{i=1}^{N}\vec{S}_{i}\vec{B},
\end{equation}
where $J>0$ is the exchange interaction constant, $H_{cr}(S_{i})$
is the crystal field acting on i-th spin ($H_{cr}=K_1S_{z}^{2}+
K_2( S_x^2- S_y^2)$,
with $z$-axis directed perpendicular to a molecule plane); for
Fe$^{3+}$ $S=5/2$. In accordance with experimental results it will
be assumed that $J\gg K$ (the typical values are $J\sim$ 10
cm$^{-1}$, $K\leq$1 cm$^{-1}$). The anisotropy in the plane($K_2$) can 
be formed artificially, e.g. by means of external electric or 
magnetic fields, pressure, or using anisotropic substrate.
Due to $C_n$-symmetry there is also 
"$n$-gonal anisotropy in the plane"($n=6(10)$
for Fe$_{6(10)}$). Following analysis can be easily reformulated 
to take into account this anisotropy. 
Let suppose that the magnetic
field is $\vec{B}=\left(0,\,0,\,B_{z}\right )$, where $B_{z}$
depends on time generally. We decompose as usually for
antiferromagnets the local spins into the two subsystems
(``sublattices'') and  it is reasonable to assume the coordinates
of two magnetic sublattices, describing the macroscopic quantum
behavior of  the antiferromagnetic molecule as degrees of freedom.
In order to describe such system  the coherent quantum states will
be used \cite{8}
\begin{equation}\label{thetphi}
  |\theta_{1},\,\varphi_{1};\,\theta_{2},\,\varphi_{2}\rangle ,
\end{equation}
where $\theta_{i},\,\varphi_{i}$ are the polar and azimuthal
angles of the spin of the magnetic sublattices. (These angles are
measured from the $z$ and $x$ axes respectively). The Lagrangian
of the system can be represented as:
\begin{eqnarray}
L&=&-\frac{M}{\gamma}\sum_{i=1}^{2}
\cos\theta_{i}\dot{\varphi}_{i}+B_{z}M\sum_{i=1}^{2}
\cos\theta_{i}  
-\frac{1}{2}K_u\sum_{i=1}^{2}\sin^2\theta_i 
-\frac{1}{2}K_{\bot}\sum_{i=1}^{2}\sin^{2}
\theta_{i}\sin^{2}\varphi_{i}-NJS^{2}
\Big[\cos\left(\theta_{1}+\theta_{2}\right)\nonumber\\
&&+\sin\theta_{1}\sin\theta_{2}
\big(1+\cos\left(\varphi_{1}-\varphi_{2}\right)\big)\Big],
\label{lagr1}
\end{eqnarray}
where $M=1/2g\mu_{B}SN,\,\,\gamma
=g\mu{B}/\hbar,\,\,K_{u}=NK_1S^{2},\,  K_{\bot}= 1/2 N K_2 S^2$.

This Lagrangian can be derived by means of the standard technique
of coherent quantum states (see for example \cite{10}). All terms
in (\ref{lagr1}) are of apparent physical meaning. The first term
is called Wess-Zumino term (kinetic energy), the second 
term is Zeeman energy, the third and fourth term are the magnetic
anisotropy and the fifth is the exchange interaction energy. 

Furthermore it is convenient to use the new variables:
\begin{equation}\label{newvar1}
  \theta = \frac{\theta_{1}+\theta_{2}}{2},\;\;\varphi
  =\frac{\varphi_{1}+\varphi_{2}-\pi}{2},
\end{equation}
\begin{equation}\label{newvar2}
    \epsilon_{1} = \frac{\theta_{1}-\theta_{2}}{2},\;\;\epsilon_{2}
  =\frac{\varphi_{1}-\varphi_{2}+\pi}{2},
\end{equation}
Then the partition function of the system can be represented as
the functional integral in the Eucledian space ($\tau=it$).
\begin{equation}\label{stat1}
  Z=\int D\epsilon_{1}D\epsilon_{2}\int D\theta D\varphi\, e^{-\int_{0}^{\hbar\beta}d\tau
  L\left(\theta,\varphi,\epsilon_{1},\epsilon_{2}\right)}.
\end{equation}
Upon integrating (\ref{stat1}) over $\epsilon_{1}$ and
$\epsilon_{2}$ one can obtain the following effective Lagrangian
depending on the variable $\varphi(\tau)$ only:
\begin{equation}\label{lagr2}
       L=\frac{\chi_{\bot}}{2}\left(\frac{\dot{\varphi}}{\gamma}
-B_{z}\right)^{2}+K_{\bot}\,\cos^{2}\varphi.
\end{equation}
The total magnetic moment of the antiferromagnetic molecule
equals to:
\begin{eqnarray}
 M_{z}=
\left\{ \begin{array}{cc}
{\chi_{\bot}\left(B_{z}-\frac{\dot{\varphi}}
                     {\gamma}\right),} &  {B_{z}\leq B_{c},}\\
 {2M,} & {B_{z}\geq{B_{c}}}, 
\end{array} \right.
\label{magn}
\end{eqnarray}
where
\begin{equation}\label{chi}
  \chi_{\bot}=\frac{Ng^{2}\mu_{B}^{2}}{4J},\;\;B_{c}=\frac{4JS}{g\mu_{B}}.
\nonumber
\end{equation}
The following classical motion equation for variable $\varphi$ can
be put in correspondence to the Lagrangian (\ref{lagr2})\cite{10,11}:
\begin{equation}\label{eqm}
  \frac{\chi_{\bot}}{\gamma^{2}}\,\ddot{\varphi}+\frac{\alpha
  M}{\gamma}\,\dot{\varphi}+ K_{\bot}\,\sin
  2\varphi -\frac{\chi_{\bot}\dot{B}_{z}}{2\gamma}=0.
\end{equation}
Inserted here is the dissipative term $\alpha
M\dot{\varphi}/\gamma$ related to the attenuation occurence in
Landau--Lifshits equations, where $\alpha$ is the dimensionless
Hilbert constant.

For  description of the thermodynamical and
nonequilibrium properties of the molecule at finite temperature
one can utilize the partition function as follows (see for example
\cite{13})
\begin{equation}\label{stat2}
  Z=\int
  d\varphi_{0}\int_{\varphi(0)}^{\varphi(\hbar\beta)}D\varphi(t)\,e^{\frac{S_{eff}\left[\varphi(\tau)\right]}{\hbar}},
\end{equation}
where the effective action $S_{eff}$ equals to
\begin{eqnarray}
  S_{eff}\left
[\varphi(\tau)\right]&=&\int_{0}^{\hbar\beta}d\tau\,
L(\varphi,\dot{\varphi},\tau)
  +\frac{1}{2}\int_{0}^{\hbar\beta}d\tau\int_{0}^{\hbar\beta}d\tau^{\prime}\,
\alpha\left(\tau-\tau\prime\right)\,
\left[\varphi(\tau)-\varphi(\tau^{\prime})\right]^{2},
\label{Seff}
\end{eqnarray}
\begin{equation}\label{alpha}
   \alpha(\tau)=\frac{\alpha
   M}{2\pi\gamma}\frac{\left(\frac{\pi}{\hbar\beta}\right)^{2}}{\sin^{2}\left(\frac{\pi\tau}{\hbar\beta}\right)}.
  \end{equation}

Henceforward we shall restrict our consideration to the
only quantum properties (at $T=$ 0 K).
The Hamiltonian of the system for the collective variable
$\varphi$ can be obtained from the Lagrangian (\ref{lagr2}) by the
following procedure. The momentum $p$ corresponding to the
independent coordinate $\varphi$ equals to
\begin{equation}\label{p}
  p=\frac{\partial
  L}{\partial\dot{\varphi}}=\frac{\chi_{\bot}}{\gamma}\left(\frac{\dot{\varphi}}{\gamma}-B_{z}\right).
\end{equation}
By further substituting this expression to the Hamilton function
${\mathcal{H}}=p\dot{\varphi}-L$ one can obtain
\begin{equation}\label{h3}
  {\mathcal{H}}=\frac{1}{2\chi_{\bot}}\left(\gamma
  p+\chi_{\bot}B_{z}\right)^{2}-\frac{\chi_{\bot}B_{z}^{2}}{2}+U(\varphi).
\end{equation}
By performing a standard quantization technique  which consists in
definition of the operators $\hat{p}$ and $\hat{\varphi}$ by means
of the commutation rule
$\left[\hat{p},\hat{\varphi}\right]=-i\hbar$ we obtain
$\hat{p}=\frac{\hbar}{i}\frac{\partial}{\partial\varphi}$. The
magnetic momentum operator is determined as
$\hat{M}_{z}=-\gamma\hat{p}$. Substituting this expression into
the Hamilton function (\ref{h3}) yields
\begin{equation}\label{h4}
{\mathcal{H}}=
\frac{1}{2\chi_{\bot}}\left(M_{z}-\chi_{\bot}B_{z}\right)^{2}-
\frac{\chi_{\bot}B_{z}^{2}}{2}+U(\varphi).
\end{equation}
The magnetic moment of the antiferromagnetic  molecule and the
angular variable $\varphi$ are the conjugate variables:
$\left[\hat{M}_{z},\hat{\varphi}\right]=ig\mu_{B}$.

 Let's consider first
the case $B_{z}=0$. It is important that the equation (\ref{h4})
is isomorphous to the equation of particle in periodic potential,
therefore one can use the results known for this model.
The eigenstates of the Hamiltonian (\ref{h4}) are the Bloch
functions
\begin{equation}\label{PSI1}
  \Psi_{sn}(\varphi+\pi)=e^{i\pi m}\Psi_{sm}(\varphi),
\end{equation}
where $m$ is an arbitrary real number. By analogy with the term
-``charge states'' - for similar states in the theory of Josephson
effect \cite{13,14} it is possible to call the states (\ref{PSI1})
as the ``continuous spin states''. The Schroedinger
equation for the Hamiltonian (\ref{h4}) is reduced to the Mathieu
equation from the theory of  which it follows that the energy
spectrum of the Hamiltonian (\ref{h4}) has a band structure, i.e.
the eigenvalues of (\ref{h3}) $E_{s}(m)$ are  the functions
defined in the appropriate Brillouin zones. At $U(\varphi) \approx 0$ the
band structure corresponds to the approximation of free electrons.
In this case the energy spectrum is determined approximately by
parabolic function
\begin{equation}\label{par}
  E_{n}(m)\approx\frac{m^{2}\left(g\mu_{B}^{2}\right)}{2\chi_{\bot}}
\end{equation}
with the forbidden bands on the boundaries of the Brillouin zones:
$m_{B}=s\,\,\left(s=1,2,\ldots S_{max}\right)$, where $S_{max}$ is
the maximal spin of the molecule. Near the Brillouin zone boundary
the wave function can be represented as follows
\begin{equation}\label{PSI2}
  \Psi(\varphi)=u(\varphi)\,e^{-iEt},
\end{equation}
where $u(\varphi)=A_{1}e^{im\varphi}+A_{2}e^{i(m+2)\varphi}$. The
Schroedinger equation for the Hamiltonian (\ref{h3}) can be
written as
\begin{equation}\label{shr}
  u^{\prime\prime}+\left(\mu^{2}-2b^{2}\cos
  2\tilde{\varphi}\right)u=0,
\end{equation}
where 
\begin{equation}
 \mu^{2}=\frac{2\chi_{\bot}E}
 {\left(g\mu_{B}\right)^{2}}, \,
 b^2= \frac{\chi_{\bot} K_{\bot}}
 {2\left(g\mu_{B}\right)^{2}}.
\nonumber
\end{equation}
Used here is a new variable $\tilde{\varphi} =\varphi + \pi/2$.
The sign ``$\sim$'' will be further omitted. Substituting
(\ref{PSI2}) to (\ref{h4}) yields
\begin{equation}\label{expr2}
  \mu^{2}=\left(\frac{m^{2}+(m+2)^{2}}{2}\pm
  \sqrt{\frac{\big(m^{2}-(m+2)^{2}\big)^{2}}{4}+b^{4}}\right).
\end{equation}
In particular from the expression (\ref{expr2}) it is clear that
at $m_{B}=1$ the forbidden band width  equals
\begin{equation}\label{delta1}
  \Delta
  E_{g}=\frac{\left(g\mu_{B}\right)^{2}}{\chi_{\bot}}b^{2} .
\end{equation}
Equations (\ref{par}),(\ref{shr}),(\ref{expr2}) determine with
sufficient accuracy the energy spectrum in the limit of first two
Brillouin zones. The forbidden band between $s-1$ and $s$ bands
equals
\begin{equation}\label{delta2}
  \Delta
  E_{s}\approx\frac{g^{2}\mu_{B}^{2}}{2\chi_{\bot}}\left(2 b^2 \right)^{s}s^{1-s}.
\end{equation}

If $B_{z}\neq 0$ in (\ref{h4}) it is possible to consider
$\chi_{\bot}B_{z}$ as a classical gauge field. The wave functions
of the Hamiltonian (\ref{h4}) can be obtained from equation
(\ref{PSI1}) by means of apparent gauge transformation. The gauge
transformation can be presented as follows:
\begin{equation}\label{PSI3}
  \Psi_{1}(\varphi)\longrightarrow\Psi(\varphi)e^{\theta(\varphi,t)},
\end{equation}
Then
\begin{equation}\label{PSI4}
 i\hbar\dot{\Psi}_{1}=\left[\frac{1}{2\chi_{\bot}}\left(\hat{M}_{z}-g\mu_{B}
\frac{\partial\theta}{\partial\varphi}-\chi_{\bot}B_{z}\right)^{2}
-u(\varphi)-\frac{\chi_{\bot}B_{z}^{2}}{2}+\hbar\dot{\theta}\right]\Psi_1 .
\end{equation}
Assuming
$\theta(\varphi,t)=-\frac{\chi_{\bot}B_{z}\varphi}{g\mu_{B}}$, we
have (the label ``1'' of the function $\Psi$ will be further
omitted)
\begin{equation}\label{PSI5}
  i\hbar\dot{\Psi}=\left[\frac{M_{z}^{2}}{2\chi_{\bot}}-\frac{K_{\bot}}{2}
\cos2\varphi-\frac{\hbar\chi_{\bot}\dot{B}\varphi}{g\mu_{B}}-\frac{\chi_{\bot}B_{z}^{2}}{2}\right]\Psi.
\end{equation}
The last term $\frac{\hbar\chi_{\bot}\dot{B}\varphi}{g\mu_{B}}$ in
equation (\ref{PSI5}) is playing the same role as energy --
$eF\cdot x$ ($F$ characterizes the electrical field and $x$ is the
electron coordinate) in the well-known problem of the dynamics of
Bloch electron in electrical field.

Let consider the behavior of the molecule magnetization process
under the action of time-dependent magnetic field $B_{z}(t)$
varying adiabatically slowly i.e.
\begin{equation}\label{ad}
  \left|\frac{\chi_{\bot}\dot{B}_{z}}{g\mu_{B}}\right| \ll\frac{K_{\bot}}{\hbar}.
\end{equation}

To describe the dynamics of the antiferromagnetic vector of the
molecule (or ``staggered magnetization'') let's  consider a wave
packet of the Bloch functions (\ref{PSI2}). Let $\bar{m}$ and
$\bar{\varphi}$ be mean values of the generalized coordinate and
momentum of the packet and the values $\Delta m$ and $\Delta
\varphi$ ($\Delta m\cdot\Delta\varphi\sim 1$) determine the packet
width. Under influence of $\dot{B}_{z}$ which induces the
rotational torque accelerating spin precession the wave packet
moves to the boundary of Brillouin zone where Bragg reflection
takes place. Here the wave packet group velocity changes its sign
to an opposite after that a new process of the wave packet
acceleration starts.

In mathematical terms this process can be described by the
following equations:
\begin{equation}\label{m}
  \dot{\bar{m}}=-\frac{\chi_{\bot}}{g\mu_{B}}\dot{B},
\end{equation}
\begin{equation}\label{phi1}
   \dot{\bar{\varphi}}=\frac{1}{\hbar}\frac{\partial E_{s}(\bar{m})}{\partial
   \bar{m}},
\end{equation}
where $E_{s}=E_{s}(\bar{m})$ is energy spectrum of (\ref{h4}) for
the s-band. Equations (\ref{m},\ref{phi1}) must be completed by
appropriate initial conditions.


In the adiabatic case the antiferromagnetic molecule remains in the state
with the definite $s$ and all observed values  as, for example,
the magnetization (\ref{magn}), accounting (\ref{m},\ref{phi1})
are the oscillating functions of time with the frequency
\begin{equation}\label{f1}
  f_{Bloch}=\frac{\chi_{\bot}\dot{B}_{z}}{g\mu_{B}}=\frac{g\mu_{B}\dot{B}_{z}N}{4J}.
\end{equation}
These
oscillations are identical to the oscillations that electrons are
subjected in crystal or superlattice under electrical field
\cite{15} (Bloch oscillations). This macroscopic quantum effect is
essentially the macroscopic quantum coherence effect induced by
time-increasing (decreasing) magnetic field.

If the external field can be represented by the following sum
\begin{equation}\label{ext}
  B=B_{0}+at+b\sin 2\pi ft,
\end{equation}
then
it is possible to expect the resonance origin at
\begin{equation}\label{f2}
  f=f_{Bloch}
\end{equation}
and on the
frequences $f=r\cdot f_{Bloch}$ as well, where $r$ is a rational
number (Stark ladder-like resonances). 
The constant  shifting field $B_{0}$ plays here role of a
chemical potential and can be used to provide optimal conditions
for observing the considered macroscopic quantum effects.

Let's present some numerical estimations. For Fe$_{6}$
and Fe$_{10}$ $J=$20.9 cm$^{-1}$ and $J=9.6$ cm$^{-1}$,
accordingly \cite{16,17,18}. The magnetic anisotropy is of  the
"easy plane" type. The constants of anisotropy are $K=$ -0.3
cm$^{-1}$/ion (Fe$_{6}$) and $K=$ -0.1 cm$^{-1}$/ion (Fe$_{10}$).
The estimation under the formulas (\ref{ad}-\ref{f1}) gives for
Fe$_{10}$: $B_{c}=$ 97 T; $\omega_{E}=\,1.7\cdot10^{13}$ rad/s,
$f_{Bl}\approx\, 10^{5}$ Hz at $\dot{B}_{z}=10^{6}$ T/s. 

Let's
examine briefly a role of dissipation. From equation (\ref{eqm})
it follows, that the relaxation time of $\varphi$ angular
perturbations equals to $\tau=(\alpha\omega_{E})^{-1}$. For
observation of the Bloch oscillations it is necessary that
$f_{Bloch}>\tau^{-1}$ that restricts the value $\dot{B}_{z}$ from
below. For the relaxation time estimation we shall take an
advantage of the experimental data of paper \cite{19} for the
antiferromagnet MnF$_{2}$ in which the exchange interaction
constant, field $B_c$ and consequently $\omega_{E}$ are close to
that for the antiferromagnetic molecules in the question. 
In the \cite{19} the
linewidth of antiferromagnetic resonance $\Delta H$ was measured
which is linked to a Hilbert constant $\alpha$ and relaxation time
by the relation $\gamma\Delta H=\alpha\omega_{E}=1/\tau$, i.e.
$\tau=(\gamma\Delta H)^{-1}$. In agreement to \cite{19} the
temperature dependence $\Delta H(T)$ at 5 K--40 K is an average
between $T^{3}$ and $T^{4}$ with the tendency to $T^{4}$ (and even
$T^{5}$) when approaching to $T=$ 5 K. At $T=$ 5 K $\Delta H\sim $
0.1 Oe, therefore $\tau\approx\,0.5\cdot10^{-6}$ s. It is possible
to expect $\tau
>$ 10$^{-3}$ s at 0.5 K.

On the other hand the condition (\ref{ad})  restricts from
above the value $\dot{B}_{z}$ . Thus, the following inequality
should be fulfilled:
\begin{equation}\label{last}
  \frac{4J}{Ng\mu_{B}\tau}\ll\dot{B}_{z}\ll\frac{\gamma
  K_{\bot}}{\chi_{\bot}}.
\end{equation}
 Accepting $K_{2}\sim 10^{-4} J$  we shall obtain following
boundary values for an area of acceptable $\dot{B}_{z}$ values:
$ 10^{4}$ T/s and $ 10^{10}$ T/s.

In conclusion, it was shown that the external time dependent
magnetic field induces new macroscopic quantum effects in the
behavior of the antiferromagnetic nanoclusters: oscillations of
the magnetization similar to the electron Bloch oscillations in
crystal and the corresponding resonances in ac-magnetic field.
These effects are related to that of the macroscopic quantum
coherence. The analogy of antiferromagnetic nanoclusters quantum
behavior and Josephson junction was revealed as well.

Author thanks B. Barbara for discussions. This work was
partially supported by RFBR (N 99-02-17830) and INTAS (N 97-705) 

\end{document}